\documentstyle[art12,12pt,epsf]{article}

\def\mathrm{\rm}

\newcommand{\bbar}{\mbox{${\overline b}$}}
\newcommand{\bbbar}{\mbox{$b\bbar$}}

\newcommand{\effb}{\mbox{${\epsilon_b}$}}
\newcommand{\effc}{\mbox{${\epsilon_c}$}}
\newcommand{\effu}{\mbox{${\epsilon_{uds}}$}}
\newcommand{\lamb}{\mbox{${\lambda_b}$}}
\newcommand{\purb}{\mbox{${\Pi_b}$}}

\newcommand {\bea} {\begin{eqnarray}}
\newcommand {\eea} {\end{eqnarray}}
\newcommand {\beq} {\begin{equation}}
\newcommand {\eeq} {\end{equation}}

\newcommand{\RBF}{0.2176}
\newcommand{\DRBFST}{0.0033}
\newcommand{\DRBFSYS}{0.0017}
\newcommand{\DRBFRC}{0.0008}
\newcommand{\PRB}{97.2\%}
\newcommand{\EFRB}{37\%}
\newcommand{\MCLAMB}{0.47\%}

\newcommand{\EFZVTU}{2\%}
\newcommand{\EFZVTC}{15\%}
\newcommand{\EFZVTB}{50\%}
                          
\begin{document}

\abovedisplayskip 3pt plus 1pt minus 1pt
\belowdisplayskip 3pt  plus 1pt minus 1pt
\abovedisplayshortskip 2pt  plus 1pt minus 2pt
\belowdisplayshortskip 3pt  plus 1pt minus 1pt

\vspace*{-3cm}
\begin{flushright}
{\small\baselineskip 10pt SLAC--PUB--96--7170\\ 
May 1996 \\ 
(T/E)
}
\end{flushright}


\begin{center}
{\Large\bf MEASUREMENT OF {\LARGE $R_b$} AT SLD}
\footnote[1]{
 This work was supported by Department of Energy
  contracts:
  DE-FG02-91ER40676 (BU),
  DE-FG03-91ER40618 (UCSB),
  DE-FG03-92ER40689 (UCSC),
  DE-FG03-93ER40788 (CSU),
  DE-FG02-91ER40672 (Colorado),
  DE-FG02-91ER40677 (Illinois),
  DE-AC03-76SF00098 (LBL),
  DE-FG02-92ER40715 (Massachusetts),
  DE-AC02-76ER03069 (MIT),
  DE-FG06-85ER40224 (Oregon),
  DE-AC03-76SF00515 (SLAC),
  DE-FG05-91ER40627 (Tennessee),
  DE-FG02-95ER40896 (Wisconsin),
  DE-FG02-92ER40704 (Yale);
  National Science Foundation grants:
  PHY-91-13428 (UCSC),
  PHY-89-21320 (Columbia),
  PHY-92-04239 (Cincinnati),
  PHY-88-17930 (Rutgers),
  PHY-88-19316 (Vanderbilt),
  PHY-92-03212 (Washington);
  the UK Science and Engineering Research Council
  (Brunel and RAL);
  the Istituto Nazionale di Fisica Nucleare of Italy
  (Bologna, Ferrara, Frascati, Pisa, Padova, Perugia);
  and the Japan-US Cooperative Research Project on High Energy Physics
  (Nagoya, Tohoku).}

\vglue 0.3cm
{\large Erez Etzion\\}
{\sl Brunel University \\
Uxbridge, Middlesex UB8 3PH\\ 
United Kingdom.}
\vglue 0.3 cm
Representing the SLD Collaboration
\footnote[2]{SLD Collboration members and institutions are listed after the
              references.}\\
{\sl Stanford Linear Accelerator Center, Stanford University \\
Stanford, California 94309, USA.}\\
\vspace {.3cm}
{\bf Abstract}
\end{center}
\vspace{3mm}
\begin{small}
\parbox{25pt}{\hspace{25pt}}
\parbox{350pt}{\noindent We report a new measurement of
$R_b=\Gamma_{Z^o\rightarrow b\overline{b}}/
\Gamma_{Z^0\rightarrow hadrons}$ using a double tag technique where  
the $b$ selection is based on topological reconstruction of 
the mass of 
the $B$-decay vertex.
The measurement was performed using a sample of 
150k hadronic $Z^0$ events collected with the
SLD at the SLAC Linear Collider during the years 1993-1995. 
The method utilizes the 3-D vertexing
abilities of the SLD CCD pixel vertex detector and
the small stable SLC beams to obtain a high $b$ tagging
efficiency of ${\bf\EFRB}$ for a purity of ${\bf\PRB}$.
The  high purity reduces the systematics introduced by charm 
contamination
and correlations with $R_c$. 
We obtain a result of
$R_b=\RBF\pm\DRBFST_{stat.}\pm\DRBFSYS_{syst.}\pm\DRBFRC_{R_c}$.
}
\end{small}

\vglue .4cm
\begin{center}
Presented at the XXXIst Rencontres de Moriond 
Electroweak Interactions and Unified Theories,\\
Les Arcs, Savoie, France, March 16-23, 1996. \\
\end{center}

\pagebreak
\baselineskip=14pt

\section{Introduction}

The fraction $R_b$ of $Z^0\rightarrow b\overline b$ events in the hadronic 
$Z^0$ decays is of special
interest in the Standard Model (SM). 
Since this is a ratio between two hadronic rates,
uncertainties from the  unknown 
oblique or QCD corrections mostly 
cancel. Therefore given the mass of the top 
(measured by CDF and D0~\cite{cdftop})
it provides through the $Zbb$ vertex radiative corrections 
a  sensitive environment to detect 
a signal for physics beyond the SM. 
The LEP and SLD measurements on variety of $Z^0$ coupling parameters
have provided  precise confirmations
of the SM predictions. 
Hence the current average value of $R_b$ 
measurements~\cite{MIKE}, which is more than 3$\sigma$ higher 
than the SM expectation, is very 
valuable window in the electroweak tests of the SM.

Our $\bbbar$ event selection utilizes a double tag technique where one 
attempts to identify separately the two
$B$ hadrons in the  event. 
It allows measurement of both 
the $R_b$ value and the efficiency for identifying a $b$ decay directly
from the data. Most recent precise LEP~\cite{RB-lifetime} and 
SLD~\cite{rbprl1,epspap} $R_b$ measurements have exploited
 the long lifetime of the
$B$-hadrons to distinguish between the $b$ and the charm or lighter 
quark events. 
The elimination of charm is
critical in obtaining precision $R_b$ measurements due to the
dominance of charm decay modeling uncertainties in the
overall $R_b$ uncertainty.
LEP measurement are already systematically limited mainly by the charm
contamination. Hence a new $b$ tag
technology is required to improve the current level of precision.
To increase significantly both the efficiency and the purity of $\bbbar$
identification our new measurement
uses the CCD pixel vertex detector 
(VXD) to reconstruct the mass of the secondary vertex.
 In this paper we will show that using
this mass tag we have obtained an $R_b$ measurement with the
best total systematic uncertainty of all current $R_b$ measurements, and
will, with data from future SLD runs, become the most precise
single measurement of $R_b$.

\section {SLD Detector}
The SLD detector has been described in reference~\cite{DESIGN}, and only
components important to this analysis are briefly reviewed here.
Charged particle tracking was performed using the Central Drift Chamber 
(CDC)~\cite{CDC}
surrounded by a 0.6~$T$ solenoidal magnetic field.
The vertex detector(VXD~\cite{VXD2}) is of special 
importance for this measurement. It consists of  480 charged 
coupled devices (CCDs) 
with 2 hits in the angular region of
$\cos \theta < 0.74$ and 1 hit within $\cos \theta < 0.8$. 
Each CCD is an array of $385 \times 578$ square pixels 
of size $22\mu m \times 22 \mu m$. The CCDs are arranged in four concentric 
layers at a radii from $29.5~mm$ to $41.5mm$ from the beam line. 
A typical tracks produces hits in two or three of these layers.
SLC provides SLD with a small  and very stable interaction point (IP)
($<rms>_{xyz}\approx 2.4\times 0.8\times 700~\mu m^3$). 
$\sigma^{IP}_{xy}$ 
measured with 
reconstructed tracks from $\sim$30 sequential hadronic $Z^0$ events is 
$7 \pm 2~\mu m$  and the $z$
position measured on event-by-event basis is 
$\sigma_z^{IP}=38~\mu m$.
The impact parameter resolution in plane perpendicular to 
(containing) the beam axis is 
$\sigma_{r \phi}[\mu m]=11\oplus 70/p \sin^{3/2} \theta$  
 ($\sigma_{rz}[\mu m]=37\oplus 70 /p \sin^{3/2} \theta $) where $p$ is in 
GeV/c. 
The energy deposition in the  Liquid Argon Calorimeter (LAC)~\cite{LAC} 
was used in the event trigger and in the calculation of the event thrust axis.

\section{Analysis Method}
\subsection{Topological Vertexing}

SLD has already presented a measurement of $R_b$  
using the now standard lifetime 
double tag methods~\cite{epspap}.
The current analysis is performed in a similar manner except the
$b$-tag on the track impact parameter to the interaction point 
is replaced by a $b$-tag using the reconstructed
mass of the secondary vertex.

The identification of the  vertices   is performed  using a topological
vertexing procedure~\cite{DJNIM}. It searches for 3-D high track 
overlapping density  location from the single track probability (resolution) 
function. 
An event is divided in to two hemispheres with the axis defined
by the 
highest momentum jet.
The vertex finding for a hemisphere is done using only tracks
within the hemisphere, and the measured IP.
A secondary (+tertiary) vertex is found in $45\%$ ($5\%$) of the $b$ 
hemispheres.
The {\it seed vertex} (SV) is defined as the 
most significant non-primary vertex~\cite{DJNIM}.
Hence, a SV is identified in $\EFZVTB$  of the $b$ hemispheres 
(and in $\EFZVTC$ and $\EFZVTU$ of the  charm and the light quark  
hemispheres respectively).
The tracks from a $b$ decay chain do not originate from a common vertex 
and will not necessarily be associated with the SV.
All unassociated tracks
are checked for consistency with the SV,
looking at  the point of closest approach
with respect to the vertex flight direction.

\begin{figure}[h] \centering
\begin{minipage}{14.0cm}
\epsfxsize=12cm
\epsfysize=11cm
\epsfbox{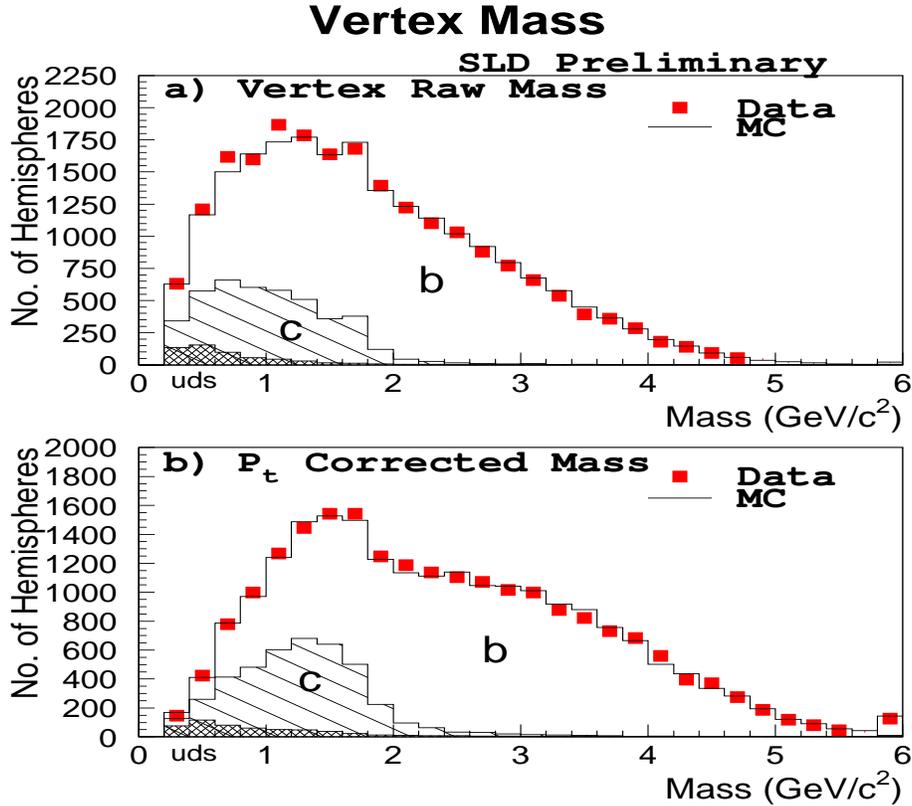}
\end{minipage}
\caption{\label{massdis}
\leftskip 1.5 cm 
\rightskip 1.5 cm
Vertex mass distribution (a), and the $P_t$ corrected mass distribution (b). 
The data is plotted with boxes where MC $b$, $c$, and $uds$ are represented by the open, hatched and cross hatched histograms respectively.}
\end{figure}

To obtain the vertex mass, all tracks associated with the SV
and consistent with it are assigned the mass of a pion and
used to calculate the invariant vertex mass. 
The mass distribution in our data compared to Monte Carlo (MC)
is shown in Fig.~\ref{massdis}a.

\textwidth     17.0cm 
\textwidth     19.5cm

\subsection{The Mass Tag}

The MC reconstructed mass distributions 
show that a sharp cut-off just above the charm mass exists, beyond which
almost only $b$ decays are found (see Fig.~\ref{massdis}a).
However since we are using charged track information to reconstruct 
the $B$ mass, 
only  half of the reconstructed $b$ masses are beyond the natural 
charm edge.
One can still improve the $b$ tagging performance by including
additional kinematic information to compensate for the loss of neutral 
particles energy
information. Comparison between the direction of the SV
displacement from the
{\it primary vertex} (PV) and the direction of the sum of 
momenta of the associated 
charged tracks 
results in the missing transverse momentum.
Including the transverse momentum $P_t$  as the  minimum missing 
momentum 
we can define  our  tagging parameter ${\cal M}$ to be:
\beq
{\cal M}\equiv\sqrt{P_t^2+M(tracks)^2}  +|P_t|\leq M_B
\eeq

This procedure increases
mass for the $b$ hemispheres significantly, especially affects those $B$s with
small charged tracks invariant mass, while charm decays close to the full charm
mass will gain relatively little.
However, the   errors in
the derivation of either the PV or the SV may cause some low mass
charm events or  $uds$ events 
to gain enough $P_t$ to enter to  the selected sample. 
Therefore, in order to prevent  fluctuations in the $P_t$ distribution
to contaminate our sample  the following constraints are added:
The contribution that is consistent with coming from the errors
on one of the two vertices is subtracted from the missing $P_t$.
The new ${\cal M}$  is limited to less than
twice  the  original mass derived from  charged tracks. 
The improvement of the tagging performance is demonstrated in  Fig.~\ref{massdis}b.

\subsection{$R_b$ Measurement}
After partitioning events into hemispheres, $R_b$ is measured from $F_s$,
the rate at which hemispheres
pass the mass tag cut (single tags), and $F_d$, the rate at which both
hemispheres in an event pass the mass tag cut (double tags).
\bea
\label{fsfd}
F_s & = &\effb R_b + \effc R_c + \effu(1-R_b-R_c), \\ \nonumber
F_d & = &(\effb^2 + \lamb(\effb-\effb^2))R_b + \effc^2 R_c + \effu^2(1-R_c-R_b).
\eea
Estimations for the light quarks and  charm
tagging rates ($\effu,\effc$) and  the 
hemisphere $b$-tagging efficiency correlation ($\lamb$) are 
derived from our Monte Carlo, where for $R_c$ we assume the SM value. 
Measuring the
two tagging rates allows $\effb$, the hemisphere $b$-tagging efficiency, 
to be calculated 
directly from data simultaneously with $R_b$ itself:

\bea
R_b &= &\frac{(F_s-R_c(\effc-\effu)-\effu)^2}
{F_d-R_c(\effc-\effu)^2+\effu^2-2F_s \effu\lamb R_b(\effb-\effb^2)}, 
\\ \nonumber
\effb &= &\frac{F_d-R_c \effc(\effc-\effu)-F_s \effu-\lamb R_b(\effb-\effb^2)}
{F_s-R_c(\effc-\effu)-\effu},
\eea
thus eliminating the
influence of uncertainties in $b$ decay modeling except in
the correlation term $\lamb$.
The correlation which expresses the difference between the 
efficiency of tagging the two hemispheres in $b$ event ($\effb^{double}$)
 and $\effb^2$ is given by:
\beq
\lamb = \frac{\effb^{double}-\effb^2}{\effb-\effb^2}.
\eeq


\subsection{$b$ Tag Performance}

\begin{figure}[tbh] \centering
\begin{minipage}{13.0cm}
\epsfxsize=13.0cm
\epsfysize=10.0cm
\epsfbox{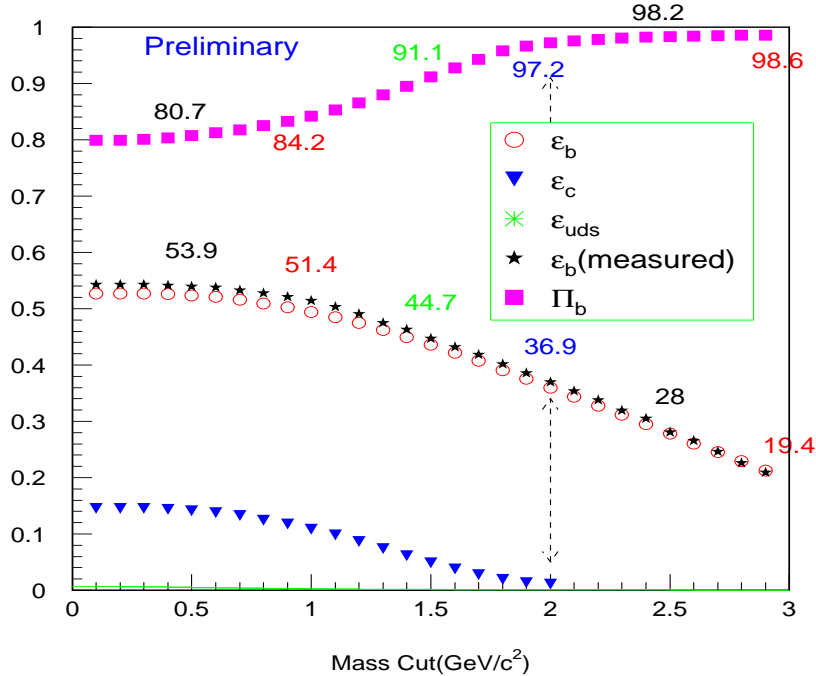}
\end{minipage}
\caption{\label{perf} 
\leftskip 1.5 cm 
\rightskip 1.5 cm
As a function of the ${\cal M}$ cut the  purity of the 
$b$ sample ($\Pi_b$) and the efficiency for tagging light flavours
($\epsilon_c$, $\epsilon_{uds})$ are shown together with the 
$b$-tag efficiency ($\effb$) as measured in the data (black stars) 
and as estimated from MC
(open circles).}
\end{figure}

The $b$ hemisphere  tagging efficiency and purity 
as a function of the cut on   ${\cal M}$ 
are shown in Fig.~\ref{perf} along with the
efficiencies estimated for charm and $uds$ hemispheres. 
At a mass cut of 2~$GeV$ a $b$-tag efficiency of $\EFRB$
and a $b$ purity of
 $\PRB$ is achieved. The measured $\effb$ from the data agrees with the
MC estimate reasonably well.
This far exceeds the performance we previously 
obtained using the impact parameter double tag
 ($\effb^{lifetime}=31\%$,
$\purb^{lifetime}=94\%$)~\cite{epspap}.

From the single and double hemisphere $b$ tagging efficiencies we obtain
the $b$ hemisphere efficiency correlation 
$\lambda_b=\MCLAMB$.

A total of 71000 events passing the standard SLD hadronic  events 
selection (see e.g.~\cite{epspap}) were included in
this analysis to obtain $R_b=\RBF\pm\DRBFST_{stat.}$
The $R_b$ measurement   is performed also with different mass cut values and the
variation in $R_b$ is found to be consistent with statitics.

\subsection{Systematic Uncertainties}

\begin{figure}[tbh] \centering
\begin{minipage}{13.0cm}
\epsfxsize=15.0cm
\epsfysize=10.0cm
\epsfbox{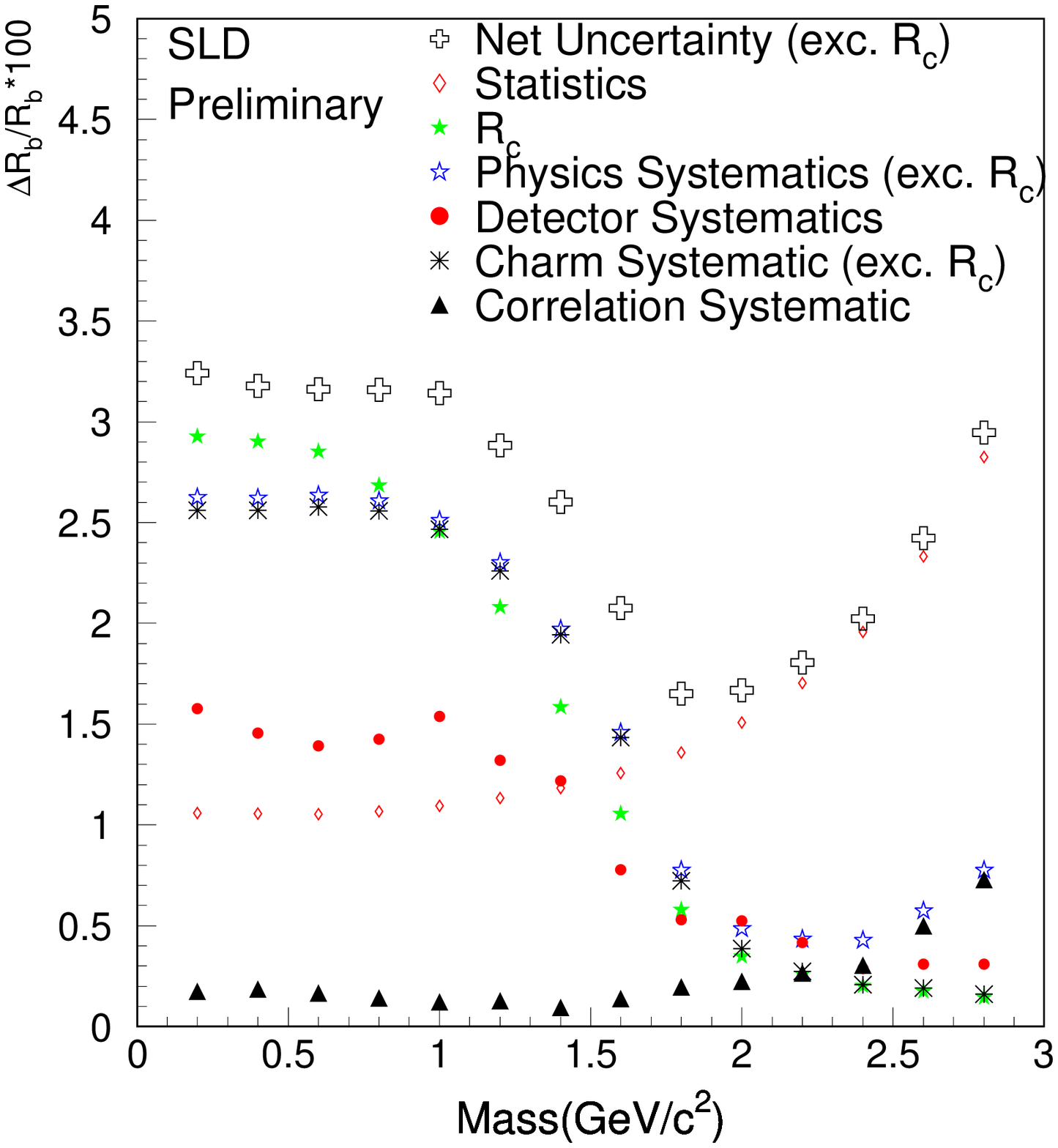}
\end{minipage}
\caption{\label{r2bsyspic} 
\leftskip 1.5 cm 
\rightskip 1.5 cm
$R_b$ statistical and systematic uncertainties versus ${\cal M}$ cut.}
\end{figure}

The systematic uncertainty is a combination of detector related effects 
such as tracking efficiency and resolution, 
as well as physics effect while the effect of $R_c$ uncertainty is treated 
separately.
The physics systematic studies are similar to those of previous $R_b$ 
measurements at LEP and SLD~\cite{RB-lifetime,epspap}
 and the error is a combination of uncertainties
from the
estimation of the correlations and the modeling of the charm and $uds$.
The curves in Fig.~\ref{r2bsyspic} show all the detector, 
physics and $R_c$ systematic and the statistical uncertainties  
versus the ${\cal M}$ 
cut. 	
It demonstrates how charm systematics dominate for a loose ${\cal M}$ cut,
 while after the natural charm mass cut-off the statistical 
uncertainty is  the primary limitation.  
The contributions to the systematic uncertainty at the optimal cut are 
summarized in Table~\ref{rbsys}. At this cut the combined uncertainty from all
systematic sources including detector, physics and $R_c$ is $\delta R_b/R_b = 0.83\%$. 

\begin{table}[tbh]
\begin{center}
\begin{tabular}{| l | c || l | c |} \hline
\multicolumn{4}{| c |}{\bf Detector Systematics} \\ \hline
Systematic & $\delta R_b/R_b$  & Systematic & $\delta R_b/R_b$  \\
\hline
Efficiency Corrections & 0.21\% &     &                \\
Z Impact Resolutions & 0.48\% & Beam Position Tails  & 0.08\%               \\
\hline
\multicolumn{4}{| l | } {Total Detector Systematics 0.53\%} \\ \hline
\multicolumn{4}{| c |}{\bf Physics Systematics} \\ \hline
Systematic   & $\delta R_b/R_b$ & Systematic             & $\delta R_b/R_b$ \\
\hline
Correlation Systematics & 0.37\% & Charm Systematics & 0.36\% \\
Light Quark Systematics & 0.19\%         & $R_c=0.171\pm0.014$ & 0.34\%            \\
\hline
\multicolumn{4}{| l |} {Total Physics (excluding $R_c$) Systematics 0.55\%} \\ \hline
\end{tabular}
\caption{\label{rbsys}
 Summary of contributions to the systematic error at ${\cal M_{cut}}=2.0~GeV$}
\end{center}
\end{table}

\section{Conclusions}
The current world average $R_b$ measurement is more than 3$\sigma$ higher 
than the SM expectation value. 
Previous LEP and SLD measurements were generally 
systematically limited mainly by the charm systematic uncertainty.
High resolution topological vertexing and precision 
knowledge of the SLD interaction point allow a new high purity 
$b$-tag to provide a low systematic approach for precision $R_b$ measurement.
Analyzing 1993 to 1995 data,  SLD has measured a new preliminary  $R_b$ value:
\bea
R_b=\RBF\pm\DRBFST_{stat.}\pm\DRBFSYS_{syst.}\pm\DRBFRC_{R_c}
\nonumber
\eea
This value supersedes our previous $R_b$ measurement.
With a new vertex detector and more data SLD 
is expected  to perform a measurement of $R_b$ to a precision of
$<1\%$. 
\pagebreak
 
\bibliographystyle{plain}

\begin{thebibliography}{60}
{\leftskip 1.2 cm 
\rightskip 1.2 cm
\bibitem{MIKE} M.~ Hildreth {\sl ``The current Status of the $R_b$ and $R_c$ 
                             puzzle''},
               {proceeding of XXXIst Rencontres de Moriond, Les Arcs,
                Savoie, France, March 16-23, 1996}.



 \bibitem{cdftop} CDF Collab. F. Abe {\it et.~al}, 
                  {\sl Phys. Rev. Lett.} {\bf{74}} (1995)  2626; \\
             D0 Collab. S.~Abachi {\it et.~al.}, 
            {\sl Phys. Rev. Lett.} {\bf{74}} (1995) 2637; \\
            updated this conference by  R. Hall (D0) and F. Tartarelli (CDF), 
            proceeding of XXXIst Rencontres de Moriond, Les Arcs,
                Savoie, France, March 16-23, 1996.


 \bibitem{RB-lifetime} ALEPH Collab. D.~Buskulic {\it et.~al},
                  {\sl Phys. Lett.} {\bf{B313}} (1993) 535; \\
                       OPAL Collab. P.~D.~Acton {\it et.~al},
                  {\sl Z. Phys.} {\bf{C60}} (1993) 579; \\
                       OPAL Collab. D.~Akers {\it et.~al},
                  {\sl Z. Phys.} {\bf{C65}} (1994) 17; \\
                       DELPHI Collab. P.~Abreu {\it et.~al},
                  {\sl Z. Phys.} {\bf{C65}} (1995) 555.

\bibitem{rbprl1}
                 SLD Collab. K.~Abe {\it et.~al.},
                  {\sl Phys. Rev.} {\bf{D53}}, (1996) 1023.

 \bibitem{epspap} SLD Collab. K.~Abe {\it et.~al.},
                 {\sl ``The Lifetime Probability Tag Measurement of $R_b$
                    using the SLD''},
                  SLAC-PUB-95-7004, 
              proceedings of the {\sl International Europhysics 
              Conference on High 
              Energy Physics}, Brussels, Belgium, July 1995. 

\bibitem{DESIGN} G.~Agnew {\it et.~al.}, 
                 {\sl SLD Design Report}, SLAC-0273 (1984).

\bibitem{CDC}  M. Fero {\it et.~al.}, 
          {\sl Nucl.   Inst. \& Meth. } {\bf A367} (1995) 111.

\bibitem{VXD2} G.~Agnew {\it et.~al.},  SLAC-PUB-5906; C.J.S. Damerell 
          {\it et.~al.} in proceedings of the {\sl 26th International Conference on High Energy Physics}, Dallas (1992) vol. 2 p. 1862.

\bibitem{LAC} D.~Axon {\it et.~al.}, {\sl Nucl. Inst. \& Meth.} {\bf A238} (1993) 
472.

\bibitem{DJNIM} D.~Jackson 
                {\sl ``A Topological Vertex Reconstruction Algorithm 
                  for Hadronic Jets''}, 
                  to be submitted to {\sl Nucl. Inst. \& Meth.}











}
\end{thebibliography}

\null
\begin{center}
{\bf The SLD Collboration}\\
\vskip .7\baselineskip

\small
                          
%
%
%
  \def\iADEL{$^{(1)}$}
  \def\iBOL{$^{(2)}$}
  \def\iBU{$^{(3)}$}
  \def\iBRUN{$^{(4)}$}
  \def\iUCSB{$^{(5)}$}
  \def\iUCSC{$^{(6)}$}
  \def\iCIN{$^{(7)}$}
  \def\iCSU{$^{(8)}$}
  \def\iCOLO{$^{(9)}$}
  \def\iCOL{$^{(10)}$}
  \def\iFER{$^{(11)}$}
  \def\iFRA{$^{(12)}$}
  \def\iILL{$^{(13)}$}
  \def\iLBL{$^{(14)}$}
  \def\iMIT{$^{(15)}$}
  \def\iMASS{$^{(16)}$}
  \def\iMISS{$^{(17)}$}
  \def\iMOSC{$^{(18)}$}
  \def\iNAG{$^{(19)}$}
  \def\iOREG{$^{(20)}$}
  \def\iPAD{$^{(21)}$}
  \def\iPERU{$^{(22)}$}
  \def\iPISA{$^{(23)}$}
  \def\iRUT{$^{(24)}$}
  \def\iRAL{$^{(25)}$}
  \def\iSOGANG{$^{(26)}$}
  \def\iSLAC{$^{(27)}$}
  \def\iTENN{$^{(28)}$}
  \def\iTOH{$^{(29)}$}
  \def\iVAND{$^{(30)}$}
  \def\iWASH{$^{(31)}$}
  \def\iWISC{$^{(32)}$}
  \def\iYALE{$^{(33)}$}
  \def\dead{$^{\dag}$}
  \def\andgen{$^{(a)}$}
  \def\andper{$^{(b)}$}
%
%
$^*$
\mbox{K. Abe                 \unskip,\iNAG}
\mbox{K. Abe                 \unskip,\iTOH}
\mbox{I. Abt                 \unskip,\iILL}
\mbox{T. Akagi               \unskip,\iSLAC}
\mbox{N.J. Allen             \unskip,\iBRUN}
\mbox{W.W. Ash               \unskip,\iSLAC$^\dagger$}
\mbox{D. Aston               \unskip,\iSLAC}
\mbox{K.G. Baird             \unskip,\iRUT}
\mbox{C. Baltay              \unskip,\iYALE}
\mbox{H.R. Band              \unskip,\iWISC
% ...
}
\it
  \vskip \baselineskip                   
  \vskip \baselineskip                   
%
%
%
  \iADEL
     Adelphi University,
     Garden City, New York 11530 \break
  \iBOL
     INFN Sezione di Bologna,
     I-40126 Bologna, Italy \break
  \iBU
     Boston University,
     Boston, Massachusetts 02215 \break
  \iBRUN
     Brunel University,
     Uxbridge, Middlesex UB8 3PH, United Kingdom \break
  \iUCSB
     University of California at Santa Barbara,
     Santa Barbara, California 93106 \break
  \iUCSC
     University of California at Santa Cruz,
     Santa Cruz, California 95064 \break
  \iCIN
     University of Cincinnati,
     Cincinnati, Ohio 45221 \break
  \iCSU
     Colorado State University,
     Fort Collins, Colorado 80523 \break
  \iCOLO
     University of Colorado,
     Boulder, Colorado 80309 \break
  \iCOL
     Columbia University,
     New York, New York 10027 \break
  \iFER
     INFN Sezione di Ferrara and Universit\`a di Ferrara,
     I-44100 Ferrara, Italy \break
  \iFRA
     INFN  Lab. Nazionali di Frascati,
     I-00044 Frascati, Italy \break
  \iILL
     University of Illinois,
     Urbana, Illinois 61801 \break
  \iLBL
     Lawrence Berkeley Laboratory, University of California,
     Berkeley, California 94720 \break
  \iMIT
     Massachusetts Institute of Technology,
     Cambridge, Massachusetts 02139 \break
  \iMASS
     University of Massachusetts,
     Amherst, Massachusetts 01003 \break
  \iMISS
     University of Mississippi,
     University, Mississippi  38677 \break
  \iNAG
     Nagoya University,
     Chikusa-ku, Nagoya 464 Japan  \break
  \iOREG
     University of Oregon,
     Eugene, Oregon 97403 \break
  \iPAD
     INFN Sezione di Padova and Universit\`a di Padova,
     I-35100 Padova, Italy \break
  \iPERU
     INFN Sezione di Perugia and Universit\`a di Perugia,
     I-06100 Perugia, Italy \break
  \iPISA
     INFN Sezione di Pisa and Universit\`a di Pisa,
     I-56100 Pisa, Italy \break
  \iRUT
     Rutgers University,
     Piscataway, New Jersey 08855 \break
  \iRAL
     Rutherford Appleton Laboratory,
     Chilton, Didcot, Oxon OX11 0QX United Kingdom \break
  \iSOGANG
     Sogang University,
     Seoul, Korea \break
  \iSLAC
     Stanford Linear Accelerator Center, Stanford University,
     Stanford, California 94309 \break
  \iTENN
     University of Tennessee,
     Knoxville, Tennessee 37996 \break
  \iTOH
     Tohoku University,
     Sendai 980 Japan \break
  \iVAND
     Vanderbilt University,
     Nashville, Tennessee 37235 \break
  \iWASH
     University of Washington,
     Seattle, Washington 98195 \break
  \iWISC
     University of Wisconsin,
     Madison, Wisconsin 53706 \break
  \iYALE
     Yale University,
     New Haven, Connecticut 06511 \break
  \dead
     Deceased \break
  \andgen
     Also at the Universit\`a di Genova \break
  \andper
     Also at the Universit\`a di Perugia \break
\rm
%
\end{center}

\end{document}